\documentclass[submitting]{nst}
\usepackage{graphicx}
\usepackage{subfigure}
\usepackage{subfigure,dcolumn}
\usepackage[T2A,T1]{fontenc}
\usepackage[russian,english]{babel}

\usepackage{listings}
\usepackage{amssymb}
 \usepackage{amsthm}
\usepackage{array}
\usepackage{fancyhdr}
\usepackage{graphicx}
\usepackage{hyperref}
\usepackage[hyphenbreaks]{breakurl}
\usepackage{lineno}

\begin{document}

\title{Development of the strip LGAD detector with double-end readout for future colliders}

\thanks{This work was supported in part by the National Natural Science Foundation of China under Grant 12042507, Grant 12175252, Grant 12275290, Grant 11961141014, Grant 12105298; in part by the China Postdoctoral Science Foundation under Grant 2022M722964; in part by the National Key R$\&$D Program of China under Grant 2022YFE0116900; in part by the State Key Laboratory of Particle Detection and Electronics under Grant SKLPDE-ZZ-202315.}

\author{Weiyi Sun}
\affiliation{Institute of High Energy Physics, Chinese Academy of Sciences, Beijing 100049, China}
\affiliation{University of Chinese Academy of Sciences, Beijing 100049, China}

\author{Mengzhao Li}
\email[Mengzhao Li, ]{mzli@ihep.ac.cn}
\affiliation{Institute of High Energy Physics, Chinese Academy of Sciences, Beijing 100049, China}
\affiliation{China Center of Advanced Science and Technology, Beijing 100190, China}
\author{Tianyuan Zhang}
\affiliation{Institute of High Energy Physics, Chinese Academy of Sciences, Beijing 100049, China}
\affiliation{University of Chinese Academy of Sciences, Beijing 100049, China}
\author{Mei Zhao}
\email[Mei Zhao, ]{zhaomei@ihep.ac.cn}
\affiliation{Institute of High Energy Physics, Chinese Academy of Sciences, Beijing 100049, China}
\author{Yunyun Fan}
\affiliation{Institute of High Energy Physics, Chinese Academy of Sciences, Beijing 100049, China}
\author{Shuqi Li}
\affiliation{Institute of High Energy Physics, Chinese Academy of Sciences, Beijing 100049, China}
\affiliation{University of Chinese Academy of Sciences, Beijing 100049, China}
\author{Yuan Feng}
\affiliation{Institute of High Energy Physics, Chinese Academy of Sciences, Beijing 100049, China}
\affiliation{University of Chinese Academy of Sciences, Beijing 100049, China}
\author{Xinhui Huang}
\affiliation{Institute of High Energy Physics, Chinese Academy of Sciences, Beijing 100049, China}
\affiliation{University of Chinese Academy of Sciences, Beijing 100049, China}
\author{Xuan Yang}
\affiliation{Institute of High Energy Physics, Chinese Academy of Sciences, Beijing 100049, China}
\author{Wei Wang}
\affiliation{Institute of High Energy Physics, Chinese Academy of Sciences, Beijing 100049, China}
\author{Zhijun Liang}
\email[Zhijun Liang, ]{liangzj@ihep.ac.cn}
\affiliation{Institute of High Energy Physics, Chinese Academy of Sciences, Beijing 100049, China}
\author{Yuekun Heng}
\affiliation{Institute of High Energy Physics, Chinese Academy of Sciences, Beijing 100049, China}

\begin{abstract}
 The Low-Gain Avalanche Diode (LGAD) is a new silicon detector and holds wide application prospects in particle physics experiments due to its excellent timing resolution. 
 The LGAD with a pixel size of 1.3 mm $\times$ 1.3 mm was used to construct a High Granularity Timing Detector (HGTD) in ATLAS experiments to solve the pile-up problem.
 Meanwhile, the Circular Electron Positron Collider (CEPC) also proposes detectors using the LGAD. However, pixel LGAD exhibits higher readout electronics density and cost, which somewhat limits the application of LGADs. To decrease the readout electronics density, the Institute of High Energy Physics (IHEP) of the Chinese Academy of Sciences has designed strip LGADs with larger areas. These strip LGADs are all 19 mm in length but with different widths of 1.0 mm, 0.5 mm, and 0.3 mm. 
 This article provides a detailed introduction to the design parameters of these strip LGADs and tests their electrical characteristics, including leakage current, break-down voltage, depletion capacitance, etc.
 The timing resolution and signal-to-noise ratio of the three strip LGAD sensors were investigated using beta sources test system.
 The position resolution parallel to the strip direction was tested and analyzed for the first time using a pico-second laser test system.
  Tests have demonstrated that the timing resolution of strip LGADs can reach about 37.5 ps, and position resolution parallel to the strip direction is better than 1 mm.
\end{abstract}

\keywords{LGAD; Silicon detector; Strip detector; Timing Resolution; Position resolution }

\maketitle

\section{Introduction}

Semiconductor detectors have garnered significant attention in the high-energy physics field due to their high spatial and timing resolution, rapid response, and flexible sizing\cite{GIROLAMO2014409,MOSER2009186,YANG2021165591}. For silicon detectors, the detection principle involves several processes: charge carriers, accelerated by electric fields greater than approximately 300 kV/cm, colliding with lattice atoms to generate secondary electron-hole pairs, leading to charge multiplication and thus converting into readable electrical signals for particle detection\cite{Maes1990ImpactII}. The gain in the multiplication process is defined as the ratio of the total number of electron-hole pairs generated during multiplication to that without multiplication. Avalanche Photodiodes (APDs), which undergo this multiplication process, are designed as high-gain devices to detect single or a few photons\cite{Mcintyre1972TheDO}. However, the high gain introduces several issues, including increased noise, device break-down due to high electric fields, difficulty in device segmentation, and increased leakage current in irradiated devices\cite{Sadrozinski_2018,Isidori_2021}. In charged particle detection, approximately 70 electron-hole pairs can be generated per micron of silicon, providing a larger initial number of electron-hole pairs compared to weak light detection, thereby allowing the use of lower gain to overcome the disadvantages of high gain.

Low Gain Avalanche Diode (LGAD) is a new kind of avalanche diode characterized by their low gain, typically around 10-50, as opposed to APDs and other types of photodiodes\cite{Moffat_2018,PELLEGRINI201412}. Designed for precise measurement of charged particle timing information, LGADs can achieve a time resolution of better than 40 ps. The Centro Nacional de Microelectronica (CNM), Hamamatsu K.K., Fondazione Bruno Kessler(FBK), and Institute of High Energy Physics, Chinese Academy of Sciences (IHEP) successfully developed the LGAD sensors\cite{PELLEGRINI201412,FERNANDEZMARTINEZ201693,Mandurrino2017NumericalSO,YANG2021165591,ZHAO2022166604,Li2023PerformanceOL,9945985,9739028,FAN2020164608}. Until now, LGAD has been proposed to be used in several particle physics experiments, such as the ATLAS High Granularity Timing Detector (HGTD)\cite{CERN-LHCC-2020-007,ALLAIRE2019355,Bruning_2022} and the CMS endcap timing layer (ETL)\cite{FERRERO2022166627,HARTMANN2019250}. 

 During upgrading of LHC to the High Luminosity LHC (HL-LHC), luminosity increases by an order, which brings severe pile-up, averaging about 200\cite{Aberle:2749422}. Therefore, the HGTD project was established and plans to use 1.3 mm pixel LGAD in the forward region of the ATLAS detector to perform timing measurements of charged tracks, helping in the suppression of pileup\cite{CERN-LHCC-2020-007,CARTIGLIA2015141}. However, the small pitch size brings a large readout electronics density and high total cost. Compared with the HL-LHC, the future electron collider, such as the Circular Electron Positron Collider (CEPC)\cite{osti_1764596,Apresyan_2020,Wada_2019,DtextquotesingleAmen2021,cepc2018cepc,cepc2018cepc2,fan2015possible},  will have lower luminosity and will no longer require high-granularity detectors for accurate position detecting. 
 
 CEPC is an ambitious collider project initiated by China that aims to construct a 100 km circumferential ring accelerator featuring two collision points\cite{cepc2018cepc,cepc2018cepc2}. Its primary scientific objectives include precise measurements of the Higgs boson's properties and electroweak physics. Additionally, it is expected to generate approximately \(10^{12}\) polarized Z bosons for flavor physics research. The momentum of final-state particles produced in collisions predominantly falls within the range of 0-5 GeV, making the detection system's ability to discern particles, particularly  \(K/\pi\) and \(K/proton\), critically important. Traditional methods based on Time Projection Chambers (TPC) and Drift Chambers (DC) for \(dE/dx\) and \(dN/dx\) measurements face particle identification limitations of  \(K/\pi\)   at around 1 GeV and \(K/proton\) at around 2 GeV \cite{ZHU2023167835}. To enhance the research in flavor physics, there is a demand for a high-precision Time-Of-Flight (TOF) detector to improve particle identification capabilities in the low-energy region. Additionally, acquiring positional information alongside timing data, which aids in pinpointing hits on outer calorimeters, would significantly benefit track reconstruction. 
 
To meet the requirements of the TOF detector of future electron colliders, especially for the CEPC, IHEP designed three strip-type LGAD sensors in a large area. These sensors are distinct in their dimensions, featuring various widths while maintaining a uniform length of 19 mm, thereby offering a unique blend of form and function tailored to specific experimental needs. Compared to the 1.3 mm pixel size LGAD sensor 
in the ATLAS HGTD project, strip LGAD achieves reducing the readout electronics density.

Compared to traditional silicon microstrip detectors, strip LGAD can also provide position resolution parallel to the strip direction thanks to its good timing performance.
In the ATLAS SCT project, the silicon microstrips detectors only provide high precision position resolution perpendicular to the direction of the microstrips; no position resolution capability parallel to the direction of the microstrips is provided\cite{ABDESSELAM2006642}. The SCT overlays two layers of silicon microstrips at a very small angle to achieve position resolution along the direction of the strips. In this work, we have achieved the position resolution parallel to the strip direction with a single layer of strip LGAD for the first time. This can provide the particle hit position in the direction parallel to the strip, for electromagnetic calorimeters adjacent to the TOF, without the need for additional position detectors,  thereby simplifying the overall composition and material budget of the CEPC detector system. Meanwhile, the position perpendicular to the strip direction can be obtained using the method in reference \cite{SUN2024169203} and \cite{Heller2022}.
 
This paper introduces the design of strip LGAD that can be used as a CEPC TOF detector, describes the time and position reconstruction methodology, and analyzes as well as the factors that affect the relevant operating parameters.

\section{Design and electric properties of strip LGAD}
\subsection{Design of the strip LGAD}

\begin{figure}[t]
\centering
\subfigure[]{
\includegraphics[width=\linewidth]{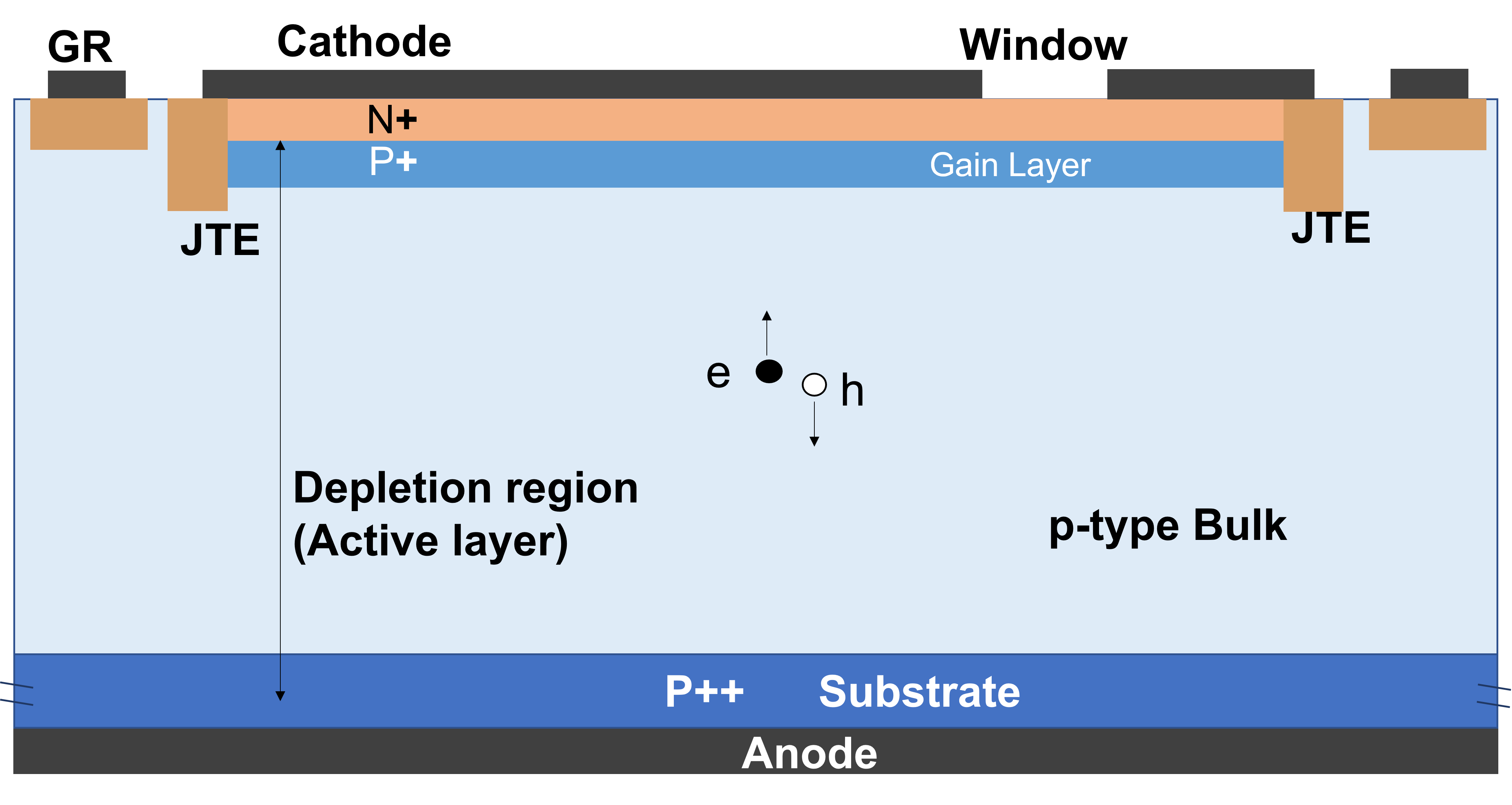}
}
\subfigure[]{
\includegraphics[width=\linewidth]{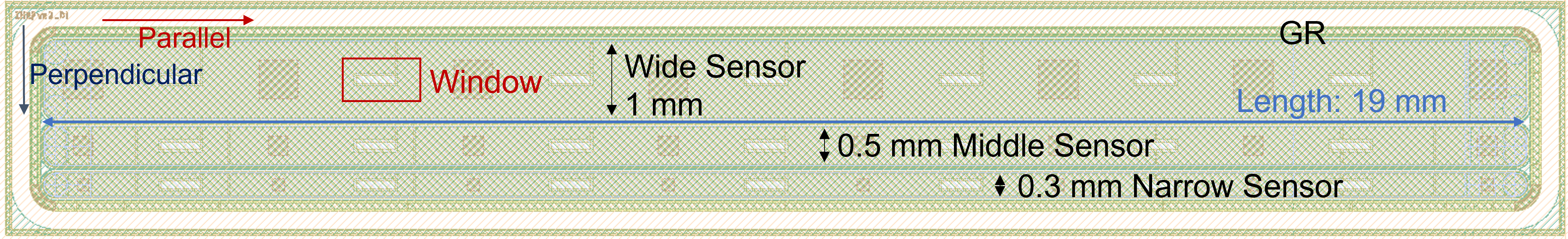}
}
\caption{(a) The structure of LGAD (not to scale), (b) the layout of the IHEP strip LGAD. The lengths of wide, middle and narrow strip sensors are all 19 mm, with widths of 1.0 mm, 0.5 mm, and 0.3 mm,  corresponding areas of 19.0 mm$^2$, 9.5 mm$^2$, and 5.7 mm$^2$ }
\label{fig:stru}
\end{figure}

The strip LGADs are fabricated on 8-inch wafers. As shown in figure \ref{fig:stru} (a), the LGAD sensors contain a high doping concentration P+ layer near the PN junction of a PIN structure, creating a strong electric field exclusively within this layer to facilitate avalanche effects, commonly referred to as the gain layer. The width and depth of this gain layer (P+ layer) are precisely manufactured through a 400 KeV ion implantation process, ranging from approximately 0.5 to 2 micrometers, designed to offer a lower gain factor (10 to 50). This design aims to amplify the signals of incident particles while maintaining low noise levels and operational stability. Additionally, to control the uniformity of the edge electric field and prevent premature device break-down, the Junction Termination Extension (JTE) and Guard Ring (GR) structures are achieved through N++ doping.

We designed three strip LGADs with a length of 19 mm, as shown in figure \ref{fig:stru} (b). The widths of these three sensors are 1.0 mm, 0.5 mm, and 0.3 mm, respectively, and are referred to as the wide, medium, and narrow sensors. The corresponding areas are 19.0 mm$^2$, 9.5 mm$^2$, and 5.7 mm$^2$, respectively. Each sensor is equipped with fully covered aluminum metal electrodes on the surface, within which seven equally spaced windows to the N+ layer are created specifically for laser testing. 

\subsection{Electric properties of the strip LGAD}

\begin{figure}
    \centering
    \includegraphics[width=\columnwidth]{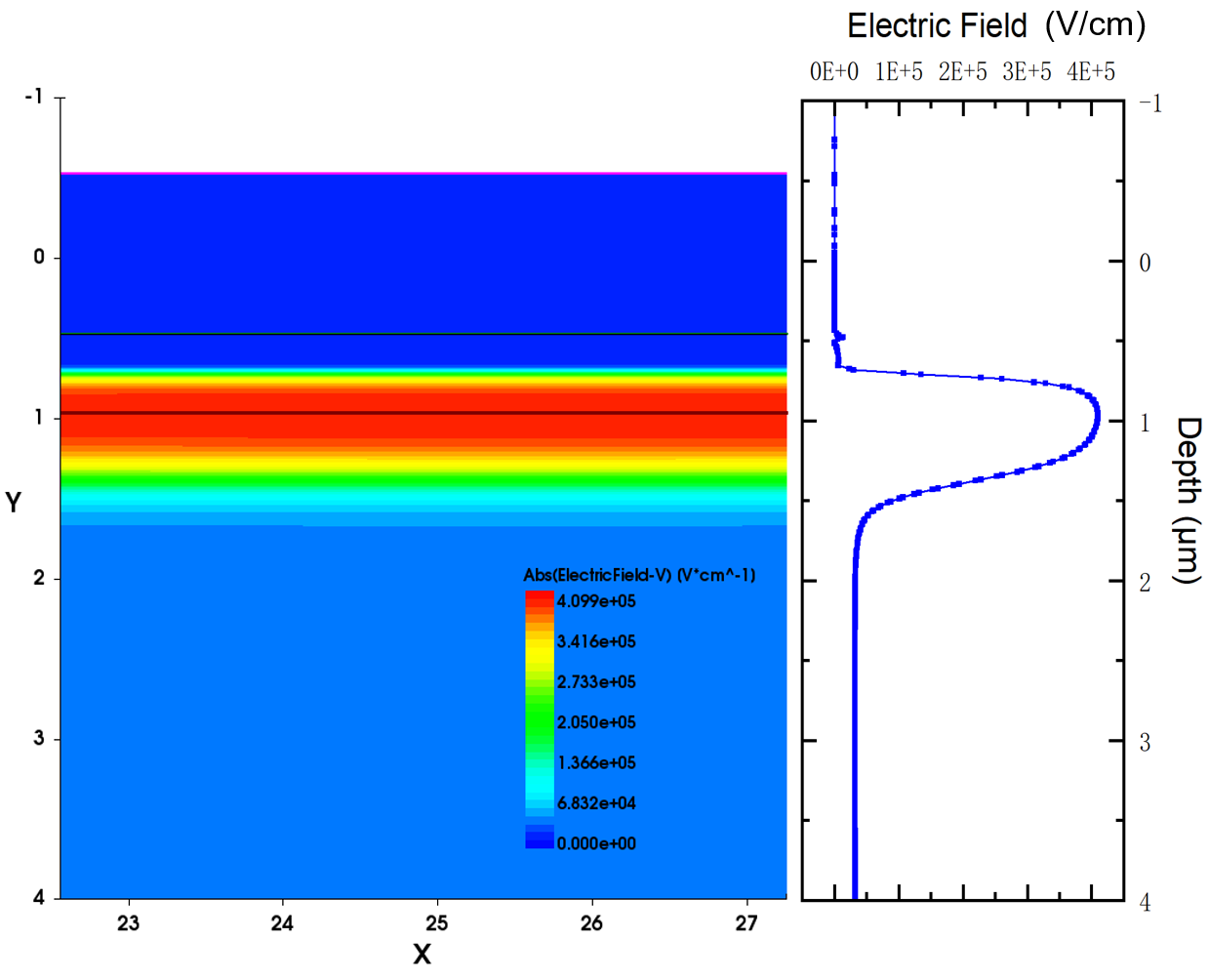}
    \caption{Electric field distribution of an LGAD in PN junction region.}
    \label{fig:EF}
\end{figure}

Figure \ref{fig:EF} presents the electric field distribution in PN junction region of an LGAD device, as determined through finite element simulation. From the figure, it is evident that the electric field within the LGAD is distinctly divided into two regions: the epitaxial layer, which has a relatively low electric field of about 30 kV/cm, allowing electrons to reach saturation drift velocity, and the gain layer, which exhibits a significantly high electric field of up to 400 kV/cm, where electron drift can lead to multiplication.

\begin{figure}
     \centering
     \includegraphics[width=\columnwidth]{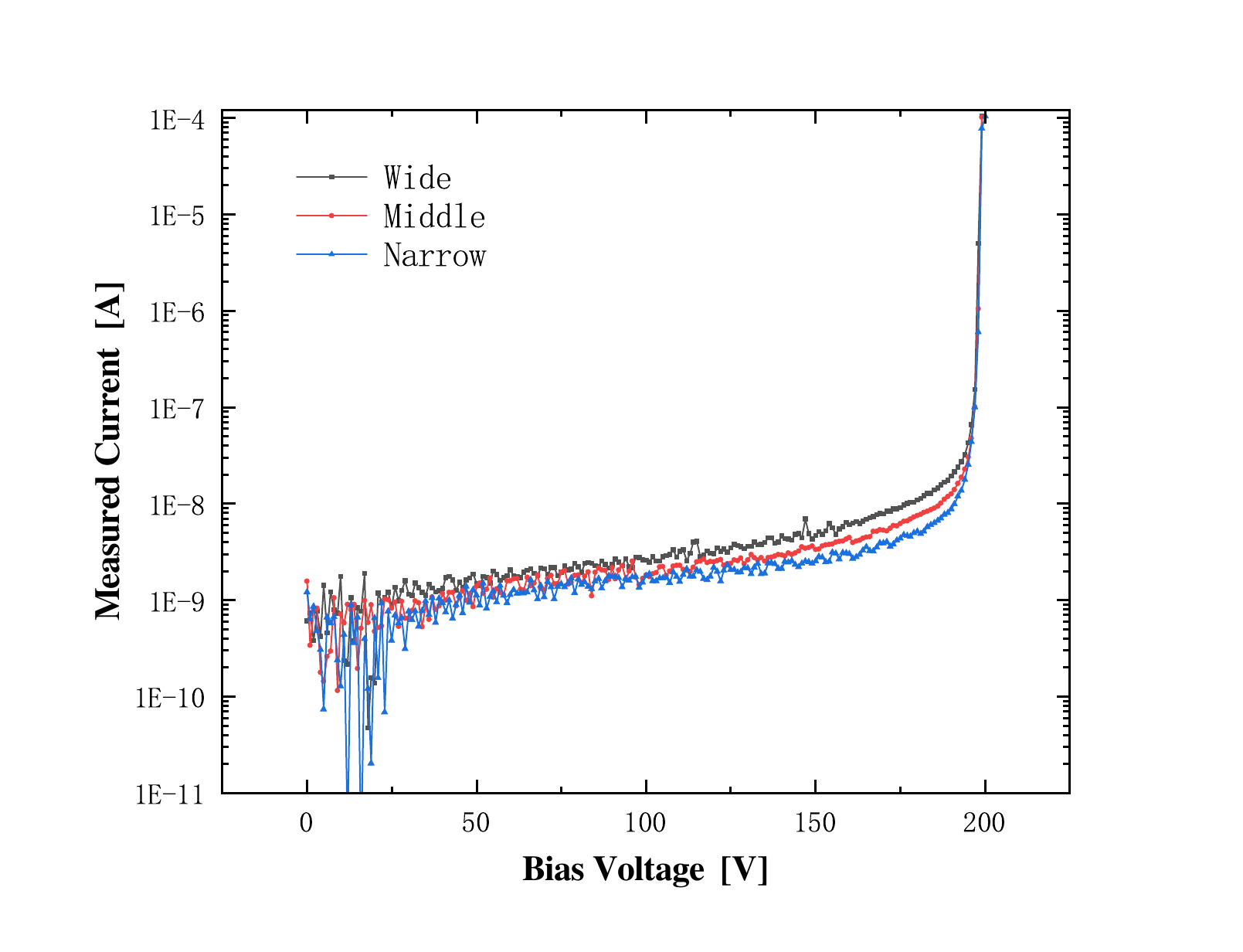}
     \caption{I-V characteristic of the strip LGADs.}
     \label{fig:IV}
 \end{figure}

\begin{figure}
     \centering
     \includegraphics[width=\columnwidth]{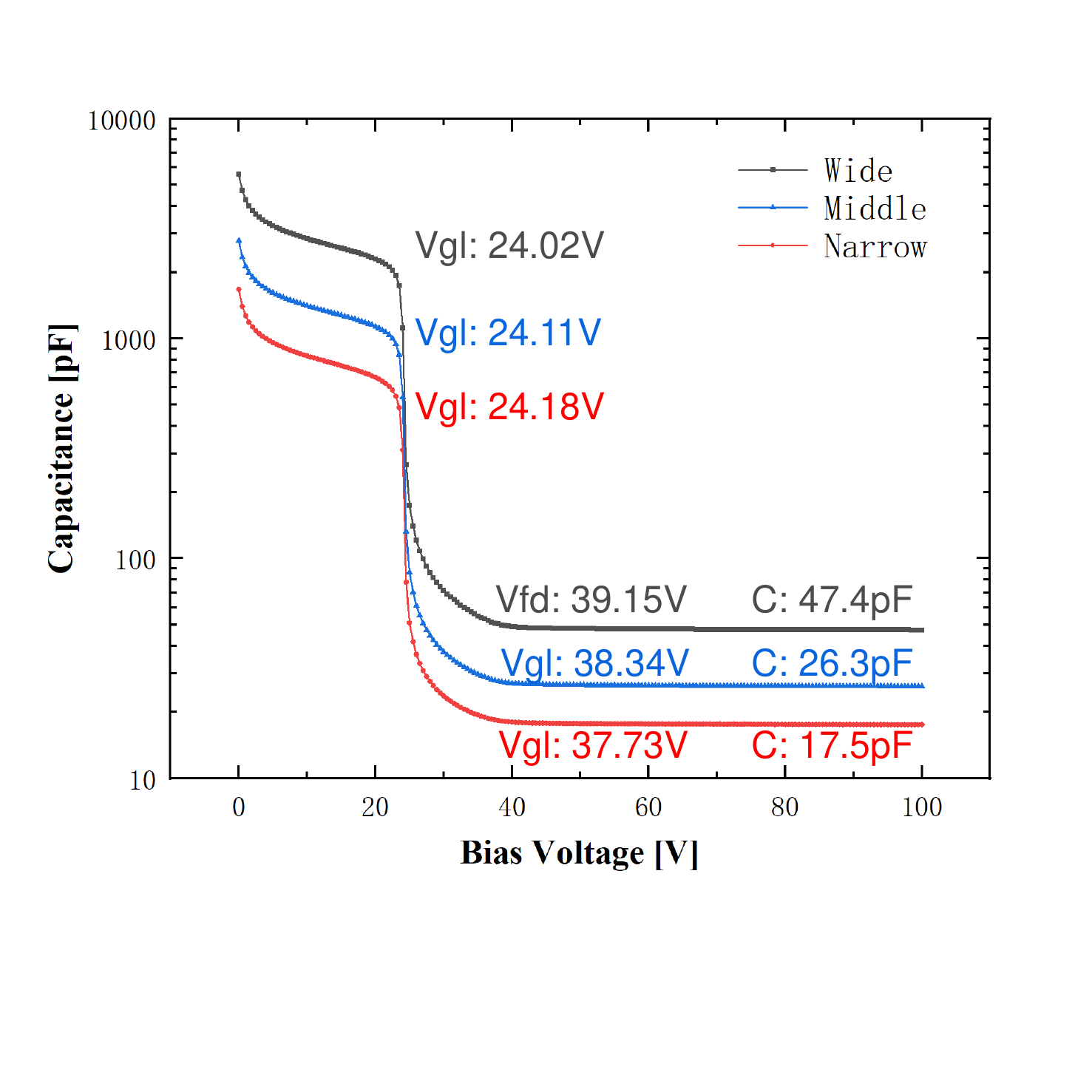}
     \caption{C-V characteristic of the strip LGADs.}
     \label{fig:CV}
 \end{figure}

The leakage current versus bias voltage (I-V) and capacitance versus bias voltage (C-V) of the three strip LGAD sensors are shown in figure \ref{fig:IV} and \ref{fig:CV}. Define bias voltage at 1 $\mu $A of leakage current as break-down voltage ($V_{BD}$) and current at 0.8 $V_{BD}$ as leakage current. Break-down voltage of the wide, middle, and narrow sensors are 198 V, 199 V, and 199V, leakage currents are 6.0 nA, 4.4 nA, and 3.0 nA. The three sensors have almost the same break-down voltage, indicating good uniformity in their manufacturing process. At the same time, the leakage current of the three sensors increases with the width (area) of the sensor. 
The C-V characteristics depicted in figure \ref{fig:CV} show three distinct phases in the capacitance variation of LGAD detectors as a function of bias voltage. Initially, at 0 V bias, the capacitance is at its highest due to the narrow depletion width between the N+ and P+ layers, exceeding 1000 pF for all devices. The next phase, marked by a gradual decrease in capacitance on the C-V curve, indicates the start of the depletion process within the gain layer. As bias voltage increases, the depletion region expands, reducing capacitance until the gain layer is fully depleted at a voltage (\(V_{gl}\)) above 24 V. After full depletion of the gain layer, depletion of the high-resistivity, low-doping epitaxial layer begins, causing a sharp decline in the C-V curve.

After the sensor is fully depleted at the full depletion voltage (\(V_{fd}\)), near 34 V, capacitance reaches its minimum and stabilizes, indicating the device's active layer full depletion status, optimal for particle detection. The depletion capacitance for wide, medium, and narrow devices are 47.4 pF, 26.3 pF, and 17.5 pF, respectively. Analysis shows that \(V_{gl}\) and \(V_{fd}\) are independent of device size, suggesting depletion behavior is mainly determined by material and structural properties, with depletion capacitance proportional to device area.

\section{Experiment setups}
\subsection{Beta source test}
To evaluate the charge collection efficiency of strip LGAD, a telescope system employing a Sr-90 beta source was established \cite{FERRERO201916,WU2023167697}, as shown in figure \ref{fig:RdBB} (b). This system facilitates the generation of electrons that traverse the LGAD detectors, triggering measurable signal responses. The strip LGAD is precisely aligned with a trigger LGAD and both ends of the strip LGAD are bonded to a four channel readout board, as shown in figure \ref{fig:RdBB} (a), and only the two outermost pads of a sensor are bonded.  Each channel of the four-channel readout board is designed based on University of California, Santa Cruz (UCSC)'s single-channel board \cite{CARTIGLIA201783}. Waveform data are collected through the double-ends readout method, meaning signals are read from both sides. The trigger LGAD, with its predefined timing resolution, serves as a timing reference for the system.
The readout board is a four-layer PCB endowed with an approximate bandwidth of 2 GHz and a trans-impedance preamplifier with a resistance of 470 $\Omega$. This setup is powered by a Keithley-2410 source meter that supplies the bias voltage to the LGAD devices, ensuring their operation in a fully depleted state, aimed at capturing the interactions between electrons and the LGADs. The readout board first amplifies the signals induced by these interactions and further amplified by a commercial amplifier with a 20 dB gain, ensuring the integrity and amplification of the signals.
Subsequently, the amplified signals are captured by an oscilloscope with a bandwidth of 2 GHz and a sampling rate of 20 Gs/s for each channel, enabling a detailed analysis of the LGADs' response to incident beta particles.

\subsection{Pico-second laser experiment}

\begin{figure}
\centering
\subfigure[]{\includegraphics[width=0.7\columnwidth]{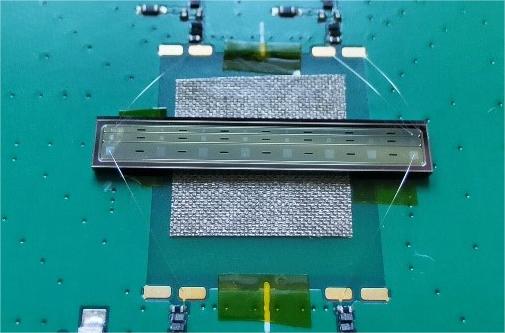}}

\subfigure[]{
\includegraphics[width=0.8\columnwidth]{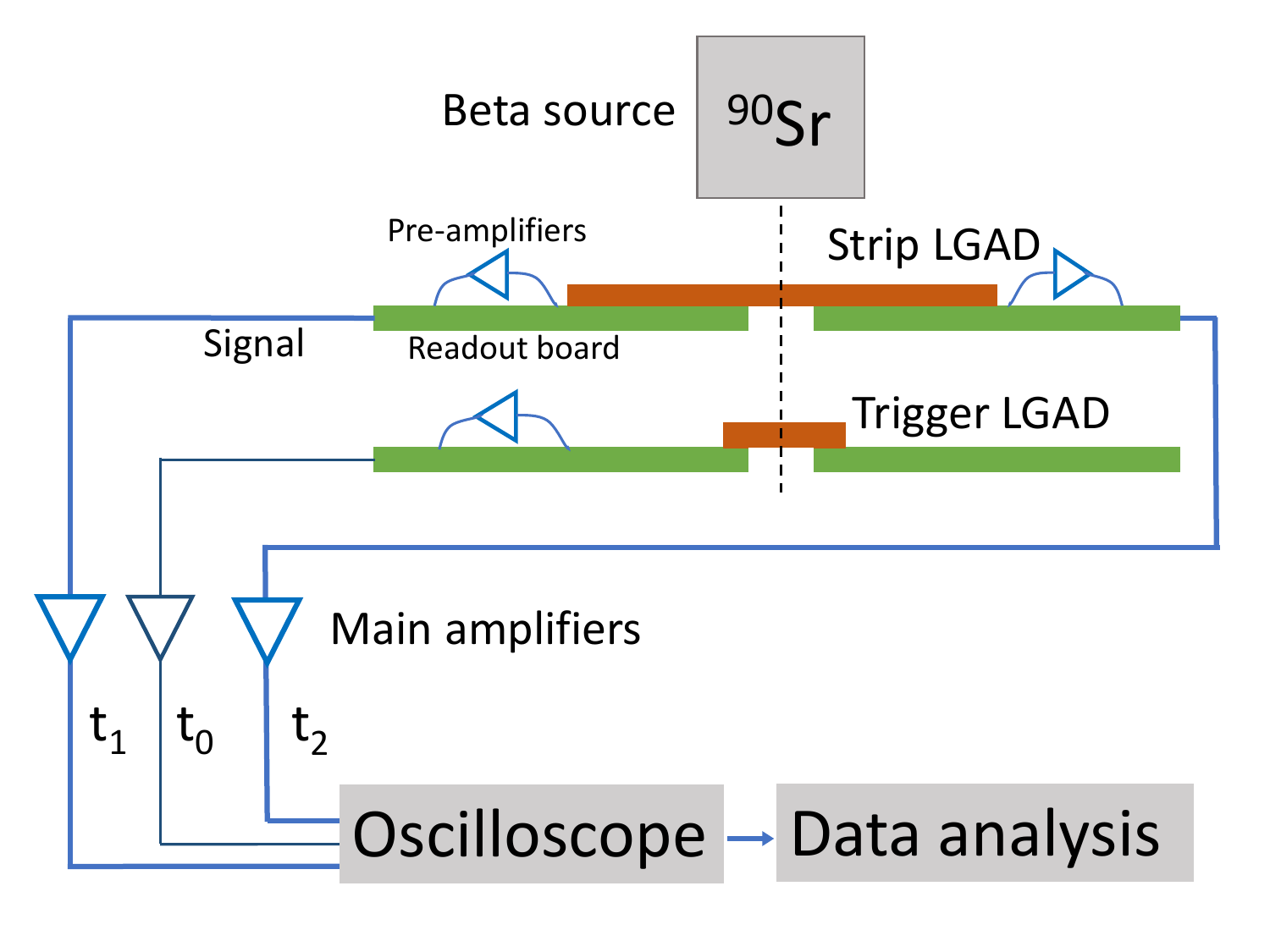}
}

\subfigure[]{
\includegraphics[width=0.8\columnwidth]{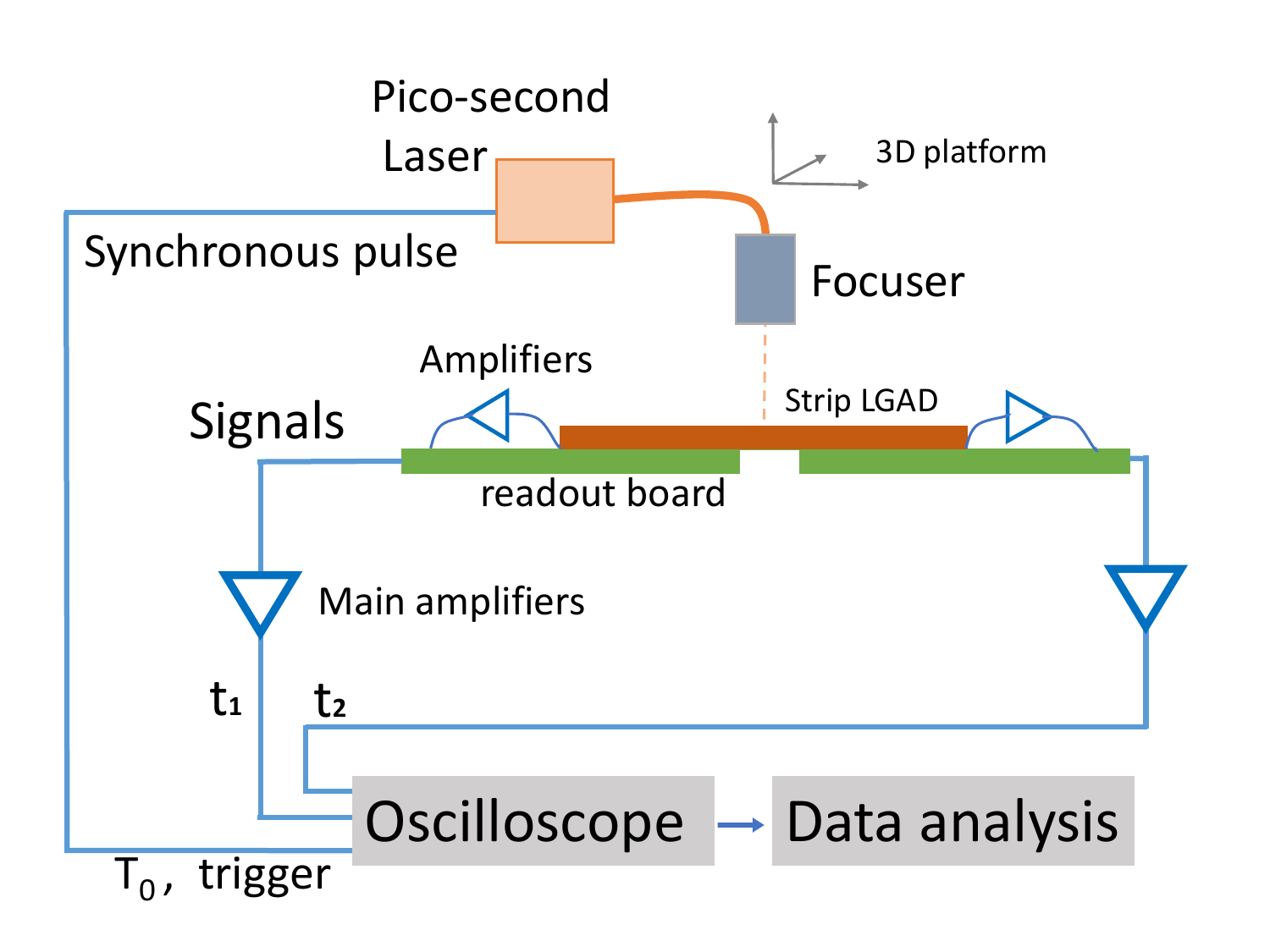}
}
\caption{(a) Both ends of the strip LGAD sensor are bonded to the readout board for double-end readout, and only the two outermost pads are bonded. (b) the telescope system of beta source test: strip LGAD sensors and trigger sensor are bonded to read-out boards and placed beneath the beta source. (c) the pico-second laser test platform, where a laser source replaces beta source. }
\label{fig:RdBB}
\end{figure}

Picosecond laser testing technology is a crucial method for investigating the position resolution of LGADs\cite{GALLRAPP201727,Medin2022}. This testing system is similar to the telescope for beta source test, but it replaces the beta source by an infrared picosecond laser source, which generates a laser pulse with a duration of approximately 7 ps and a wavelength of 1064 nm and are focused onto the strip LGAD sensors through an optical fiber and a focusing lens, generating signals, as shown in figure \ref{fig:RdBB} (c). These signals, after amplification, are recorded by an oscilloscope along with the synchronous trigger signal generated by the laser pulse for offline analysis. The laser pulse strikes the window on the metal electrode, inducing a pulse signal from the LGAD, as shown in figure \ref{fig:stru}(a). The position of the laser spot on the LGAD surface is precisely controlled using a high-precision 3D positioning stage, enabling detailed scanning of the LGAD surface and thereby facilitating a comprehensive evaluation of the device's performance. The double-end readout method provides the amplitude and time difference of peak, which is the base of position reconstruction.

\section{Results of Beta source and laser tests}\label{sec:RBT}
\subsection{Signal Property}
\begin{figure}
     \centering
     \includegraphics[width=\columnwidth]{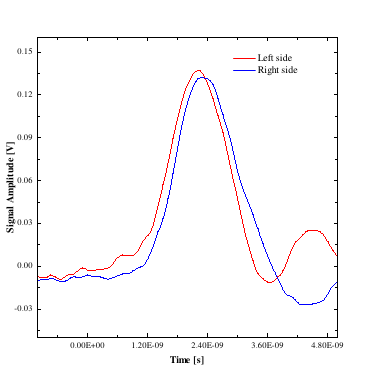}
     \caption{Waveforms read out from the left and right ends of the device upon exposure to a single laser pulse near the right end.}
     \label{fig:signal}
 \end{figure}

Figure \ref{fig:signal} displays the waveforms read out from the left and right ends of the strip LGAD's surface when exposed to a single laser hit. Due to the varying distances of the laser from the left and right readout sections, the peak amplitudes and positions of the readout signals differ. Based on the characteristics of these signals, various properties of the LGAD, such as the position resolution, can be studied.

\subsection{Timing Resolution}

 The timing resolution $\sigma_t^2$ of LGAD can be expressed as follows\cite{Sadrozinski_2018}:
\begin{equation}
    \sigma_t^2=\sigma^2_{Landau}+\sigma^2_{TimeWalk}+\sigma^2_{Jitter}+\sigma^2_{Distortion}+\sigma^2_{TDC}
\end{equation}
Each of these terms will be discussed in detail in the following.

The nature of timing resolution of LGADs is non-uniformities associated with the energy deposition of charged particles within silicon, which leads to a Landau distribution.  These deviations in energy deposition engender fluctuations in the amplitude of resultant signals, thereby introducing variability in the time measurements of incident particle events, referred to as the Landau term. The presence of the Landau term is due to the physical processes of ionization energy loss, and therefore, this term cannot be eliminated by improving the testing methods.

Furthermore, the non-uniform distribution of current signals and fluctuations in signal amplitude can induce variability in the time taken to reach a specific threshold, resulting in time walk effects, denoted by \(\sigma_{TimeWalk}\). The time walk can be eliminated by employing a constant fraction discriminator (CFD) method, which standardizes the signal processing to a constant fraction of the signal amplitude, thereby reducing the dependency on signal amplitude variations. In addition to these factors, the device-specific characteristics and electronic noise contribute to the timing jitter, represented by \(\sigma_{Jitter}\). In addition, at the rising edge of the signal, noise present either in the signal itself or in the electronic devices can cause the comparator to either trigger prematurely or delay, leading to $\sigma_{Jitter}$. This parameter is directly proportional to the system noise \( N \), and inversely proportional to the slope of the signal near the comparator threshold.

The motion of charged particles in solids is described by the Ramo-Shockley theory\cite{Shockley1938,Ramo1939}, which states that the drift velocity is directly proportional to the electric field strength but decreases in rate until saturation. Therefore, low electric field strength or non-uniform injection can affect the drift velocity of charged particles ionized at different locations within the epitaxial layer\cite{LASTOVICKAMEDIN2022167388}, subsequently impacting the readout waveform and thus affecting the time resolution. This influence is represented by the $\sigma_{distortion}$ term.  Fortunately, the strip LGAD sensors, which is designed similar to a parallel plate capacitor and maintains an electric field greater than 10 kV/cm, can establish a uniform strong field internally. Therefore, the impact of the distortion term on time resolution can be negligible.

Finally, the timing signal is converted into a digital signal by a Time-to-Digital Converter (TDC). In this converter, the leading edge of the discriminator signal is digitized and placed within a time segment of width $\Delta$T, which is determined by the least significant bit of the TDC. This time segment introduces the $\sigma_{TDC}$ term. This term can be considered negligible in the experimental setup described in this paper.

In practice, the timing resolution $\sigma_t$ of the sensor is calculated through the following equation:

\begin{equation}
\sigma_t=\sqrt{\sigma^2_{\Delta T}-\sigma^2_{Trigger}}
\end{equation}

Here $\sigma_{Trigger}$ is the timing resolution of trigger LGAD, which is 28.5 ps. $\sigma_{\Delta T}$ is the variation of flight time of a MIP between the trigger LGAD and the strip LGAD, which is defined as the sigma of the following distribution:
\begin{equation}
\Delta T=T_{trigger}-\frac{T_1+T_2}{2}
\end{equation}
$T_{1},T_{2}$ and $T_{trigger}$ are hit times measured on the right and left sides of the strip LGAD and trigger LGAD through the CFD method.

\begin{figure}[ht]
\centering
\includegraphics[width=\columnwidth]{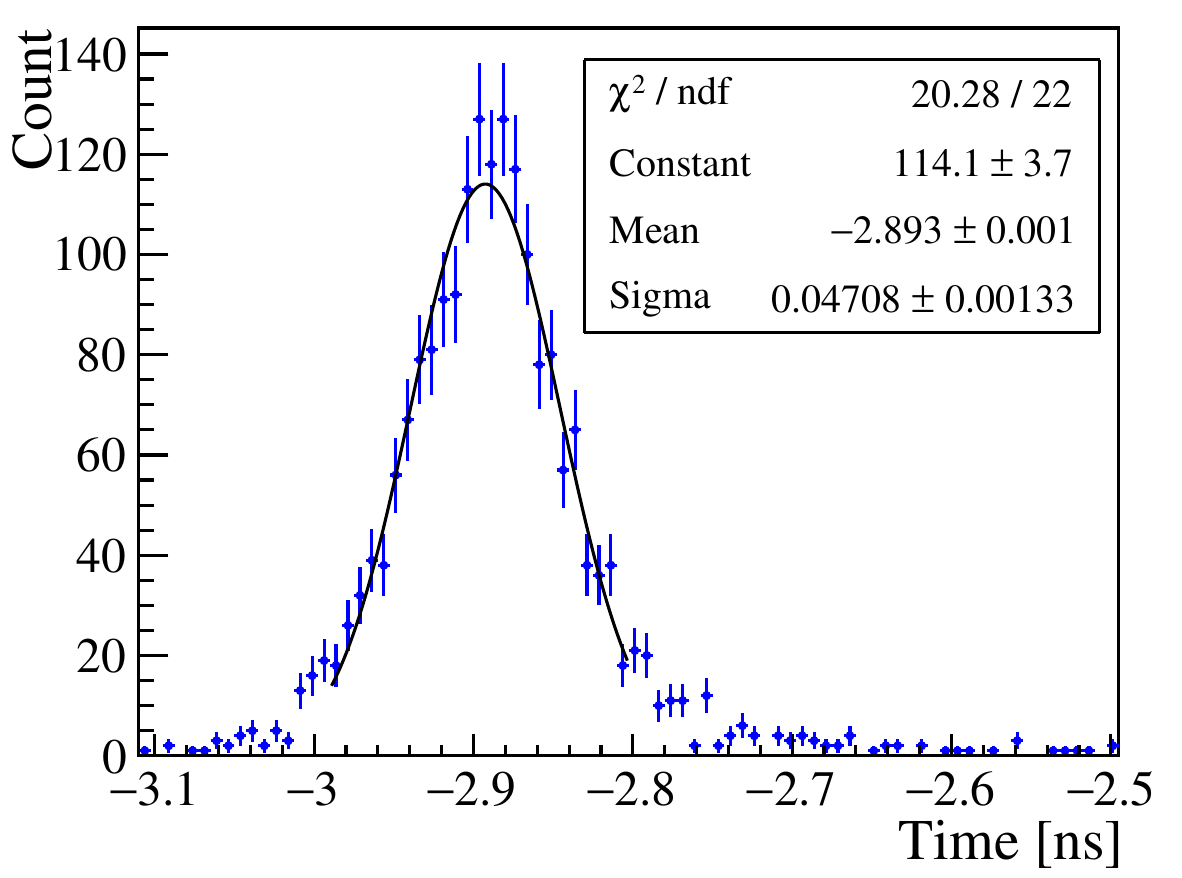}
\caption{Distribution of $\Delta T=T_{trigger}-\frac{T_1+T_2}{2}$ of narrow sensor at 185 V with gaussian fitting. The Sigma of $\sigma_{\Delta T}$ is 47.1 ps, so the timing resolution $\sigma_t$ is 37.5 ps.}
\label{fig:CH11}
\end{figure}

Figure \ref{fig:CH11} shows the distribution of $\Delta T=T_{trigger}-\frac{T_1+T_2}{2}$ of the narrow sensor at 185 V with Gaussian fitting, from which the timing resolution $\sigma_t$ is 37.5 ps.

\begin{figure}[ht]
\centering
\includegraphics[width=\columnwidth]{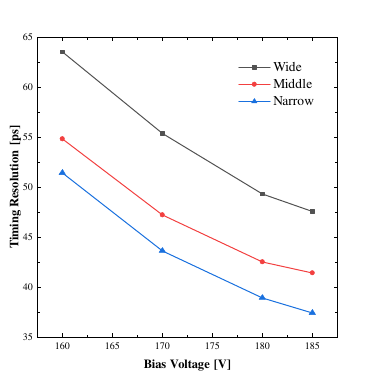}
\caption{Timing resolution of strip LGAD vs. bias voltage.}
\label{fig:TResBA}
\end{figure}

\begin{figure}[ht]
\centering
\includegraphics[width=\columnwidth]{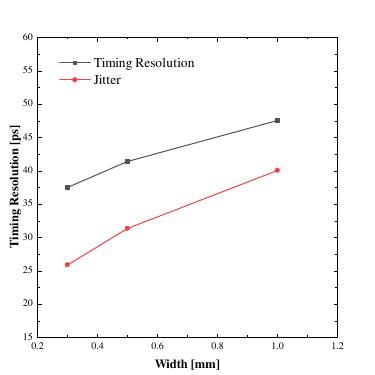}
\caption{Signal rise time vs. strip sensors' width at a bias of 185 V.}
\label{fig:TResBB}
\end{figure}

\begin{figure}[ht]
\centering
\includegraphics[width=\columnwidth]{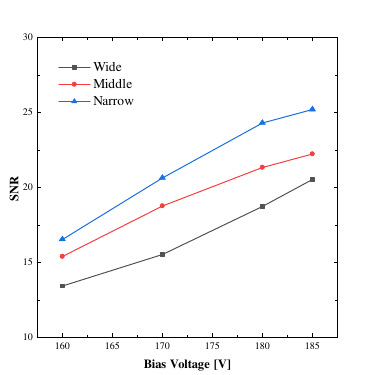}
\caption{The SNR of three strip LGAD sensors as a function of bias voltage.}
\label{fig:TResBC}
\end{figure}

\begin{figure}[ht]
\centering
\includegraphics[width=\columnwidth]{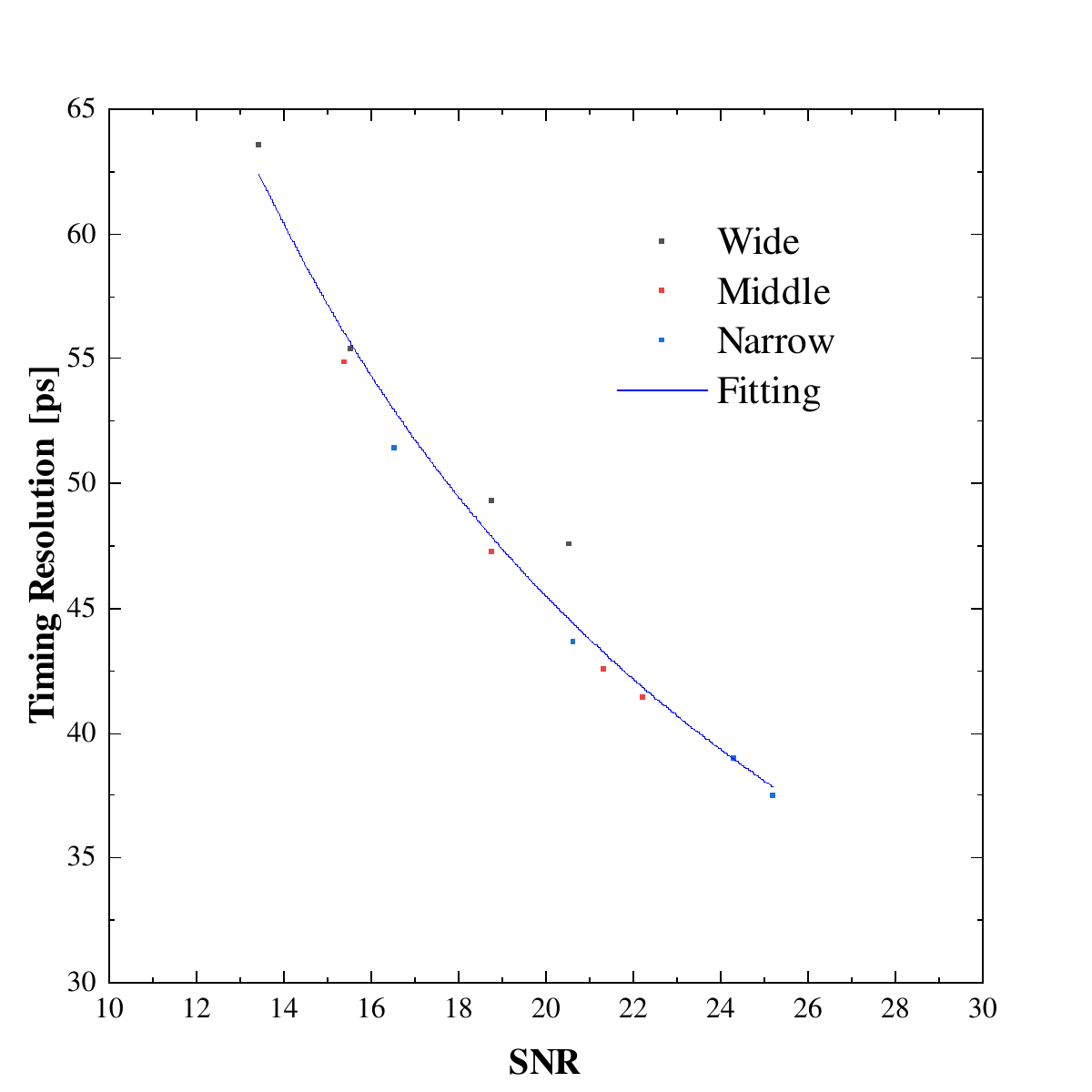}
\caption{Timing resolution vs. SNR, fitted with an inverse proportional function.}
\label{fig:TResBD}
\end{figure}

Figure \ref{fig:TResBA} shows the time resolution of three LGADs obtained from beta source testing. The best time resolutions of wide, middle, and narrow sensors are 47.5 ps, 41.4 ps, and 37.5 ps, respectively.  Figure \ref{fig:TResBB} and \ref{fig:TResBC} depict the signal rise time and signal-to-noise ratio of the three devices. The rise time refers to the duration of signal waveform amplitude increase, corresponding to the time required for electrons to drift from the furthest point in the epitaxial layer to the gain layer. Practically, the rise time is defined as the duration for the amplitude of a signal like in figure \ref{fig:signal} to escalate from 10\% to 90\% of the peak value. The rise times \(t_r\) of the three devices are 0.82 ns, 0.70 ns, and 0.65 ns, and they are found to not significantly change with voltage. The narrower device has the fastest signal rise time. Jitter has the following relationship with $t_r$ and signal-to-noise ratio (SNR):
\begin{equation}
\sigma_{jitter}=\frac{t_{r}}{SNR}    
\end{equation}
The SNR of all three devices increases with an increase in voltage, and the narrow sensor has the best SNR. As the device width decreases, the capacitance decreases, resulting in faster signal rise times and better SNR, ultimately achieving better time resolution. It should also be noted that the distance between jitter and total time resolution is almost constant as in figure \ref{fig:TResBB}, which indicates that $\sigma_{landau}$ does not vary with device area, which is beneficial for fabricating devices with larger areas. From figure \ref{fig:TResBC}, it can be seen that the SNR increases with higher voltage and improved area. Figure \ref{fig:TResBD} shows that the position resolution changes with SNR, approximating an inverse proportion relationship, meaning the value of the resolution decreases rapidly with an increase in SNR. Therefore, by slightly increasing the operating voltage and optimizing manufacturing processes and electronic methods to enhance SNR, excellent position resolution can also be achieved on large-area LGADs.

\subsection{Charge Collection}
\begin{figure}
     \centering
     \includegraphics[width=\columnwidth]{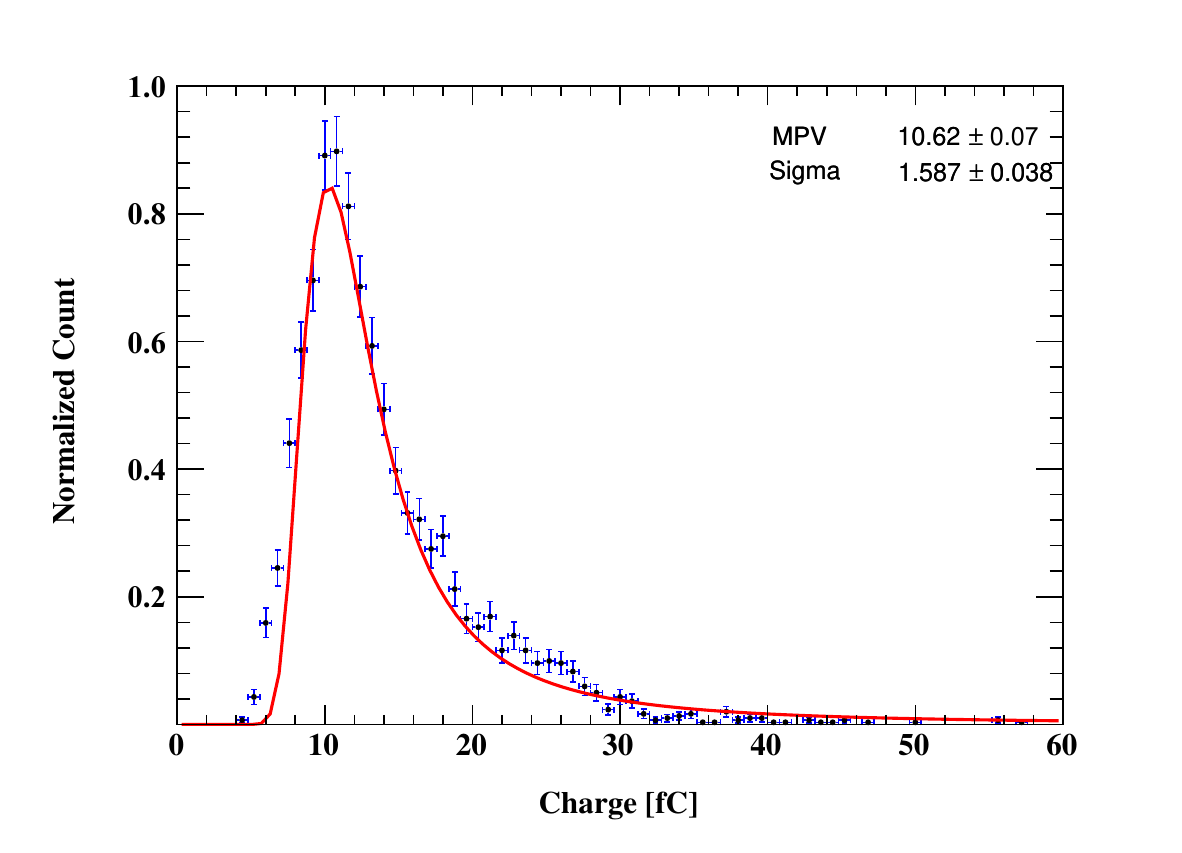}
     \caption{Distribution of collected charge of the wide sensor at 170 V. MPV of the distribution is fitted with landau function to be 10.6$\pm 0.8$ fC.}
     \label{fig:cC}
 \end{figure}

The collected charge refers to the total number of charges after the multiplication of initial electrons in the gain layer. This is typically determined by integrating the signal waveform. As shown in figure \ref{fig:cC}, the distribution of the collected charge for the middle sensor at 180V is a Landau distribution. This is because the ionization of charged particles in silicon follows a Landau distribution\cite{Meroli_2011}. Here, we define the Most Probable Value (MPV) of this distribution as the collected charge of the device at 180 V.  The collected charge of all three LGAD sensors 
increases with the increasing of bias voltage. The collected charge of all sensors exceeds 7 fC after a bias voltage greater than 160 V.  Middle sensor has the largest amount of collected charge in all bias voltages and reaches 13.7 fC at 185 V, at which point narrow and middle sensors collect 12.8 and 12.6 fC. 
 
\begin{figure}
     \centering
     \includegraphics[width=0.8\columnwidth]{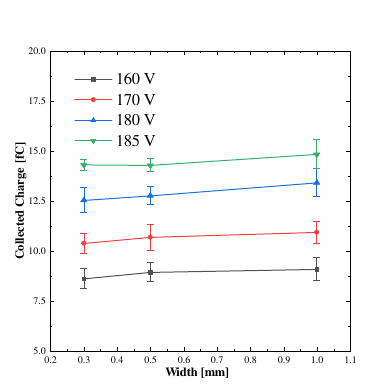}
     \caption{Collected charge of three strip LGAD sensors as a function of sensor width. The amount of charge collected basically does not change with the sensor width}
     \label{fig:CC2}
 \end{figure}

The relationship between collected charge and LGAD sensor area is shown in figure \ref{fig:CC2}. From the figure, it can be seen that the amount of charge collected basically does not change with the sensor width, which reflects the uniformity of gain and is consistent with the theory that the avalanche amplification process is only related to the thickness of the gain layer.
 
 \begin{figure}
     \centering
     \includegraphics[width=0.8\columnwidth]{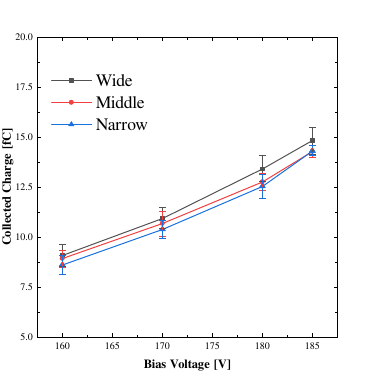}
     \caption{The collected charge of the three strip LGADs varies with the bias voltage.}
     \label{fig:CC}
 \end{figure}
 
\subsection{Position resolution}

\begin{figure}[!htb]
     \centering
     \includegraphics[width=1\columnwidth]{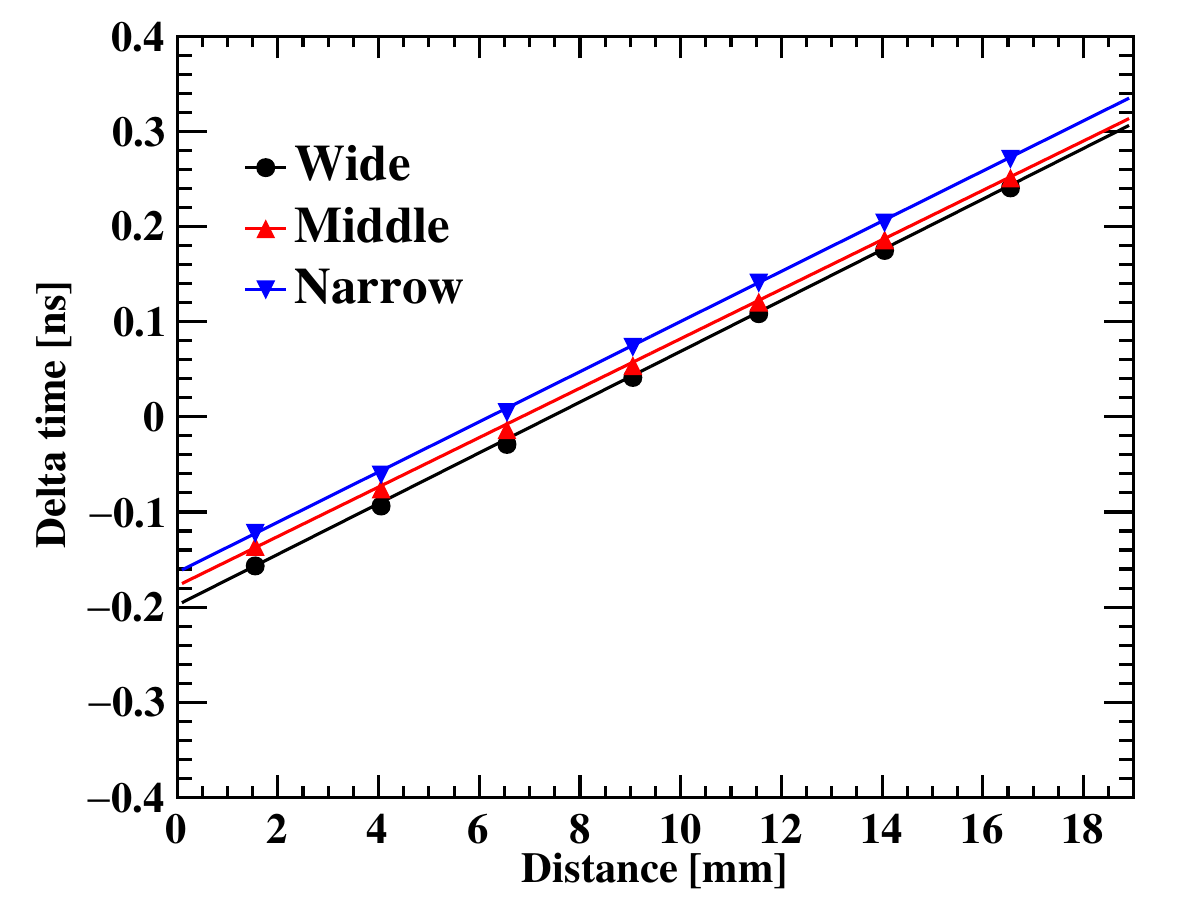}
     \caption{The delta time varies linearly with the laser hit position, which holds for three types of sensors. The position coordinates can be reconstructed by measuring delta time based on this linear relationship.}
     \label{fig:DelT}
 \end{figure}

In the detector design of CEPC, the TOF detector is located adjacent to the inner side of the electromagnetic calorimeter. The TOF detector based on strip LGAD can measure particle flight time while also providing position information parallel to the strip direction, which can assist to position reconstruction of calorimeter.
The position resolution parallel to the strip direction is studied by laser test, which the laser hits on the windows in figure \ref{fig:stru}. Since waveforms are read out from both ends, the time difference has a good linear relationship with the hit position, as shown in figure \ref{fig:DelT}. One can conclude from this figure that the slopes of the fitted lines of the three sensors are equal, with slight differences in the intercepts. The hitting position can be reconstructed through such a linear relationship.

\begin{figure}
     \centering
     \includegraphics[width=\columnwidth]{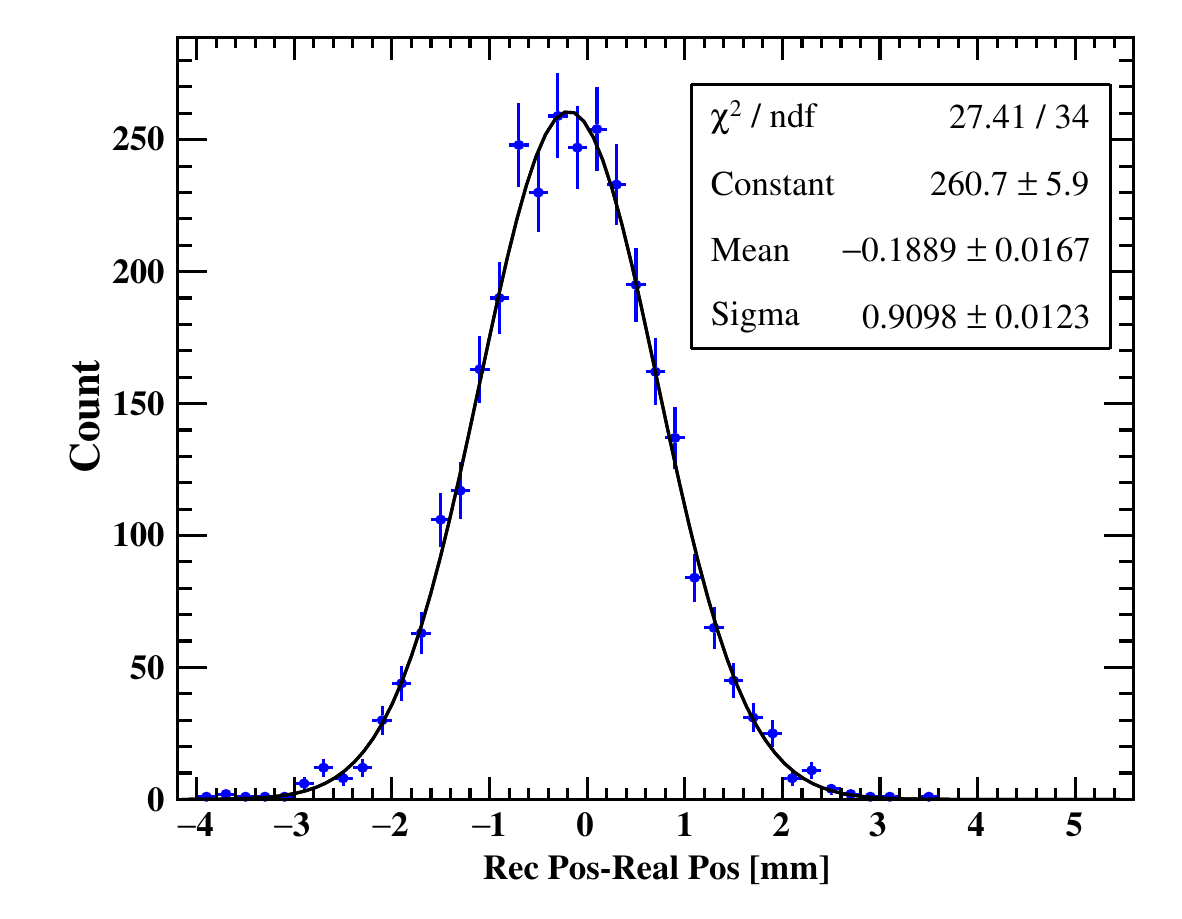}
     \caption{Distribution of the difference between the real and reconstructed position of the wide sensor window 2. The sigma of the distribution is 0.9 mm.}
     \label{fig:sres2}
 \end{figure}

\begin{figure}

\includegraphics[width=1\columnwidth]{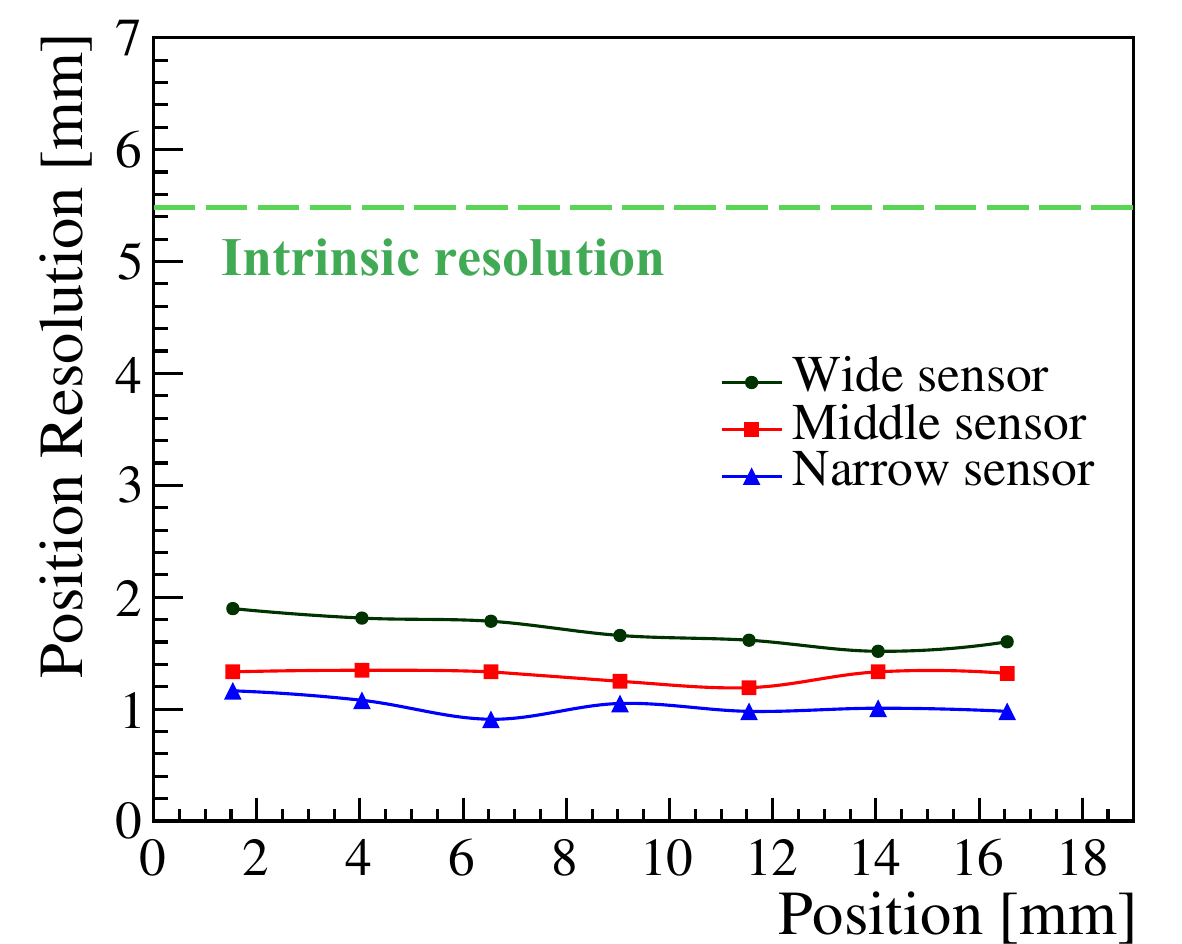}
\caption{Position resolution parallel to the strip direction as a function of test position. The intrinsic position resolution parallel to the strip direction of the strip LGAD is 5.48 mm, and the position resolution parallel to the strip direction of the narrow sensor has reached 0.9 mm.}
\label{fig:Sres}
\end{figure}

 The distribution of reconstructed points obtained from multiple laser pulses at the same point represents the position resolution of that window. For example, for the narrow sensor, the sigma of the distribution of the difference between the reconstruction and the actual position measured from the second window is 0.9 mm as shown in figure \ref{fig:sres2}, which is also defined as the position resolution of that window.
 The distribution of position resolution is shown in figure \ref{fig:Sres}.  The best position resolution is better than 0.9 mm, and the narrow LGAD has better position resolution in the direction of parallel strip. Simultaneously, as can be seen in the diagram, the intrinsic position resolution of the strip LGAD is the distance between two bonded pads divided by $\sqrt{12}$ = 5.48 mm, meaning that this method has improved the position resolution of the strip LGAD by more than five times.

Due to the good conductivity of metal cathodes, hits at different positions result in lesser differences in signal amplitude.
Therefore, it is necessary to remove the full metal coverage of the surface and place the metal cathode only at the ends and increase the resistivity of the N+ layer to improve the signal difference between the two ends and improve the positional resolution.

\section{Conclusion}\label{sec:Con}
Strip LGAD sensors provide a solution for reducing the density of readout electronics, particularly suitable for particle physics experiments with low particle flux, such as in electron collider experiments. The Institute of High Energy Physics (IHEP) has designed and manufactured three strip LGAD sensors, each 19 mm in length, with widths of 1 mm, 0.5 mm, and 0.2 mm, respectively. I-V tests show that leakage current increases with the width of the LGAD, but the break-down voltage of all LGAD sensors are all 198.7$\pm 0.47$ V. The C-V tests reveal that the capacitance of the devices increases with their width with values of 17.5 pF, 26.3 pF, and 47.4 pF for the narrow, middle, and wide. The beta source tests indicate that the charge collection quantity increases with the operating voltage, while the timing resolution improves as the device width decreases, achieving around 37.5 ps, and the jitter component of the timing resolution increases with device width. Position reconstruction parallel to the strip direction was achieved for the first time through the time difference in signal arrival at both ends, and laser tests have shown that narrower devices can achieve position resolution better than 1 mm. It was also noted that the full metal coverage on the device surface, which leads to excellent conductivity, results in minimal signal difference at both readout ends when particles strike at different positions. Therefore, in the next version, the surface metal coverage will be removed, and the resistivity of the N+ layer will be increased to enhance the signal amplitude difference at both ends of the device, thereby improving position resolution. This type of strip LGAD achieves position resolution better than 1mm in the direction parallel to the electrode and a timing resolution close to 37.5 ps, which can address the issue of \(K/\pi\) and \(K/proton\) identification failure around 1 GeV momentum for CEPC, significantly reducing the density of readout electronics and offering broad application prospects in future collider experiments.


\begin{thebibliography}{}

\end{thebibliography}


\begin{thebibliography}{99}
\bibitem{GIROLAMO2014409} B. Girolamo, M. Nessi et al., "8.23 - Silicon Trackers." In: A. Brahme (Ed.), Comprehensive Biomedical Physics, Elsevier, Oxford, 409-418 (2014). \href{https://doi.org/10.1016/B978-0-444-53632-7.00628-6}{DOI: https://doi.org/10.1016/B978-0-444-53632-7.00628-6}

\bibitem{MOSER2009186} H. Moser, "Silicon detector systems in high energy physics." Prog. Part. Nucl. Phys. 63, 186-237 (2009). \href{https://doi.org/10.1016/j.ppnp.2008.12.002}{DOI: https://doi.org/10.1016/j.ppnp.2008.12.002}

\bibitem{YANG2021165591} Y. Yang, S. Xiao, Y. Fan et al., "Characterization of the first prototype NDL Low Gain Avalanche Detectors (LGAD)." Nucl. Instrum. Meth. Phys. Res. A 1011, 165591 (2021). \href{https://doi.org/10.1016/j.nima.2021.165591}{DOI: https://doi.org/10.1016/j.nima.2021.165591}

\bibitem{Maes1990ImpactII} W. Maes, K.M. De Meyer, R. Van Overstraeten, "Impact ionization in silicon: A review and update." Solid-State Electron. 33, 705-718 (1990). \href{https://doi.org/10.1016/0038-1101(90)90183-F}{DOI: 10.1016/0038-1101(90)90183-F}

\bibitem{Mcintyre1972TheDO} R. J. Mcintyre, "The distribution of gains in uniformly multiplying avalanche photodiodes: Theory." IEEE Trans. Electron Devices 19, 703-713 (1972). \href{https://doi.org/10.1109/T-ED.1972.17485}{DOI: 10.1109/T-ED.1972.17485}

\bibitem{Sadrozinski_2018} H.F.-W. Sadrozinski, A. Seiden, N. Cartiglia, "4D tracking with ultra-fast silicon detectors." Rep. Prog. Phys. 81, 026101 (2018). \href{https://doi.org/10.1088/1361-6633/aa94d3}{DOI: 10.1088/1361-6633/aa94d3}

\bibitem{Isidori_2021} T. Isidori, P. McCavana, B. McClean et al., "Performance of a low gain avalanche detector in a medical linac and characterisation of the beam profile." Phys. Med. Biol. 66, 135002 (2021). \href{https://doi.org/10.1088/1361-6560/ac0587}{DOI: 10.1088/1361-6560/ac0587}

\bibitem{Moffat_2018} N. Moffat, R. Bates, M. Bullough et al., "Low Gain Avalanche Detectors (LGAD) for particle physics and synchrotron applications." J. Instrum. 13, C03014 (2018). \href{https://doi.org/10.1088/1748-0221/13/03/C03014}{DOI: 10.1088/1748-0221/13/03/C03014}

\bibitem{PELLEGRINI201412} G. Pellegrini, P. Fernández-Martínez, M. Baselga, et al., "Technology developments and first measurements of Low Gain Avalanche Detectors (LGAD) for high energy physics applications." Nucl. Instrum. Meth. Phys. Res. A 765, 12-16 (2014). \href{https://doi.org/10.1016/j.nima.2014.06.008}{DOI: https://doi.org/10.1016/j.nima.2014.06.008}

\bibitem{FERNANDEZMARTINEZ201693} P. Fernández-Martínez, D. Flores, S. Hidalgo et al., "Design and fabrication of an optimum peripheral region for low gain avalanche detectors." Nucl. Instrum. Meth. Phys. Res. A 821, 93-100 (2016). \href{https://doi.org/10.1016/j.nima.2016.03.049}{DOI: https://doi.org/10.1016/j.nima.2016.03.049}

\bibitem{Mandurrino2017NumericalSO} M. Mandurrino, N. Cartiglia, A. Staiano, R. Arcidiacono, M. M. Obertino, M. Ferrero, F. Cenna, V. Sola, M. Boscardin, G. Paternoster, F. Ficorella, L. Pancheri, G. F. Dalla Betta, "Numerical Simulation of Charge Multiplication in Ultra-Fast Silicon Detectors (UFSD) and Comparison with Experimental Data." In: 2017 IEEE Nuclear Science Symposium and Medical Imaging Conference (NSS/MIC), 1-4. \href{https://doi.org/10.1109/NSSMIC.2017.8532702}{DOI: 10.1109/NSSMIC.2017.8532702}


\bibitem{ZHAO2022166604} M. Zhao, X. Jia, K. Wu et al., "Low Gain Avalanche Detectors with good time resolution developed by IHEP and IME for ATLAS HGTD project." Nucl. Instrum. Meth. Phys. Res. A 1033, 166604 (2022). \href{https://doi.org/10.1016/j.nima.2022.166604}{DOI: https://doi.org/10.1016/j.nima.2022.166604}

\bibitem{Li2023PerformanceOL} M. Li, W. Sun, Z. Liang et al., "The Performance of Large-Pitch AC-LGAD With Different N+ Dose." IEEE Trans. Nucl. Sci. 70, 2134-2138 (2023). \href{https://doi.org/10.1109/TNS.2023.3289032}{DOI: 10.1109/TNS.2023.3289032}


\bibitem{9945985} F. Yuan, H. Xinhui, Y. Chengjun et al., "Study of the Acceptor Removal Effect of LGAD." IEEE Trans. Nucl. Sci. 69, 2324-2329 (2022). \href{https://doi.org/10.1109/TNS.2022.3221482}{DOI: 10.1109/TNS.2022.3221482}

\bibitem{9739028} L. Mengzhao, F. Yunyun, J. Xuewei et al., "Effects of Shallow Carbon and Deep N++ Layer on the Radiation Hardness of IHEP-IME LGAD Sensors." IEEE Trans. Nucl. Sci. 69, 1098-1103 (2022). \href{https://doi.org/10.1109/TNS.2022.3161048}{DOI: 10.1109/TNS.2022.3161048}


\bibitem{FAN2020164608} Y. Fan, S. Alderweireldt, C. Agapopoulou et al., "Radiation hardness of the low gain avalanche diodes developed by NDL and IHEP in China." Nucl. Instrum. Meth. Phys. Res. A 984, 164608 (2020). \href{https://doi.org/10.1016/j.nima.2020.164608}{DOI: 10.1016/j.nima.2020.164608}

\bibitem{CERN-LHCC-2020-007} ATLAS Collaboration, "Technical Design Report: A High-Granularity Timing Detector for the ATLAS Phase-II Upgrade," CERN, Geneva, 2020. [Online]. Available: https://cds.cern.ch/record/2719855

\bibitem{ALLAIRE2019355} C. Allaire, "A High-Granularity Timing Detector in ATLAS: Performance at the HL-LHC." Nucl. Instrum. Methods Phys. Res. A 924, 355-359 (2019). \href{https://doi.org/10.1016/j.nima.2018.05.028}{DOI: 10.1016/j.nima.2018.05.028}

\bibitem{Bruning_2022} O. Brüning, H. Gray, K. Klein et al., "The scientific potential and technological challenges of the High-Luminosity Large Hadron Collider program." Rep. Prog. Phys. 85, 046201 (2022). \href{https://doi.org/10.1088/1361-6633/ac5106}{DOI: 10.1088/1361-6633/ac5106}

\bibitem{FERRERO2022166627} M. Ferrero, "The CMS MTD Endcap Timing Layer: Precision timing with Low Gain Avalanche Diodes." Nucl. Instrum. Meth. Phys. Res. A 1032, 166627 (2022). \href{https://doi.org/10.1016/j.nima.2022.166627}{DOI: 10.1016/j.nima.2022.166627}

\bibitem{HARTMANN2019250} F. Hartmann, "Silicon-based detectors at the HL-LHC." Nucl. Instrum. Methods Phys. Res. A 924, 250-255 (2019). \href{https://doi.org/10.1016/j.nima.2018.08.101}{DOI: 10.1016/j.nima.2018.08.101}


\bibitem{Aberle:2749422} O. Aberle, I. Béjar Alonso, O. Brüning, P. Fessia, L. Rossi, L. Tavian et al., "High-Luminosity Large Hadron Collider (HL-LHC): Technical Design Report," CERN Yellow Reports: Monographs, CERN, Geneva, 2020. \href{https://doi.org/10.23731/CYRM-2020-0010}{DOI: 10.23731/CYRM-2020-0010}

\bibitem{CARTIGLIA2015141} N. Cartiglia, R. Arcidiacono, M. Baselga et al., "Design optimization of ultra-fast silicon detectors." Nucl. Instrum. Methods Phys. Res. A 796, 141-148 (2015). \href{https://doi.org/10.1016/j.nima.2015.04.025}{DOI: 10.1016/j.nima.2015.04.025}

\bibitem{osti_1764596} J. Adam, "EIC Yellow Report," 2021. \href{https://doi.org/10.2172/1764596}{DOI: 10.2172/1764596}

\bibitem{Apresyan_2020} A. Apresyan, W. Chen, G. D'Amico et al., "Measurements of an AC-LGAD strip sensor with a 120 GeV proton beam," J. Instrum. 15, P09038 (2020). \href{https://doi.org/10.1088/1748-0221/15/09/P09038}{DOI: 10.1088/1748-0221/15/09/P09038}

\bibitem{Wada_2019} S. Wada, K. Ohnaru, K. Hara et al., "Evaluation of characteristics of Hamamatsu low-gain avalanche detectors," Nucl. Instrum. Meth. A 924, 380-386 (2019). \href{https://doi.org/10.1016/j.nima.2018.09.143}{DOI: 10.1016/j.nima.2018.09.143}


\bibitem{DtextquotesingleAmen2021} G. D'Amen et al., "2021 IEEE Nuclear Science Symposium and Medical Imaging Conference (NSS/MIC)," (2021). \href{https://doi.org/10.1109/nss/mic44867.2021.9875914}{DOI: 10.1109/nss/mic44867.2021.9875914}

\bibitem{cepc2018cepc} CEPC Study Group, "CEPC conceptual design report: Volume 1-accelerator." arXiv preprint arXiv:1809.00285 (2018). \href{https://doi.org/10.48550/arXiv.1809.00285}{DOI: 10.48550/arXiv.1809.00285}

\bibitem{cepc2018cepc2} CEPC Study Group, "CEPC Conceptual Design Report: Volume 2 - Physics \& Detector." arXiv preprint arXiv:1811.10545 (2018). \href{https://doi.org/10.48550/arXiv.1811.10545}{DOI: 10.48550/arXiv.1811.10545}

\bibitem{fan2015possible} J. Fan, M. Reece, L.-T. Wang, "Possible futures of electroweak precision: ILC, FCC-ee, and CEPC." J. High Energy Phys. 2015, 196 (2015). \href{https://doi.org/10.1007/JHEP09(2015)196}{DOI: 10.1007/JHEP09(2015)196}

\bibitem{ZHU2023167835} Y. Zhu, S. Chen, H. Cui, M. Ruan, "Requirement analysis for dE/dx measurement and PID performance at the CEPC baseline detector." Nucl. Instrum. Methods Phys. Res. A 1047, 167835 (2023). \href{https://doi.org/10.1016/j.nima.2022.167835}{DOI: 10.1016/j.nima.2022.167835}

\bibitem{ABDESSELAM2006642} A. Abdesselam, T. Akimoto, P.P. Allport et al., "The barrel modules of the ATLAS semiconductor tracker." Nucl. Instrum. Methods Phys. Res. A 568, 642-671 (2006). \href{https://doi.org/10.1016/j.nima.2006.08.036}{DOI: 10.1016/j.nima.2006.08.036}

\bibitem{SUN2024169203} W. Sun, M. Li, Z. Liang et al., "The performance of AC-coupled Strip LGAD developed by IHEP." Nucl. Instrum. Methods Phys. Res. A 1062, 169203 (2024). \href{https://doi.org/10.1016/j.nima.2024.169203}{DOI: 10.1016/j.nima.2024.169203}

\bibitem{Heller2022} R. Heller, C. Madrid, A. Apresyan et al., "Characterization of BNL and HPK AC-LGAD sensors with a 120 GeV proton beam." J. Instrum. 17, P05001 (2022). \href{https://doi.org/10.1088/1748-0221/17/05/P05001}{DOI: 10.1088/1748-0221/17/05/P05001}


\bibitem{FERRERO201916} M. Ferrero, R. Arcidiacono, M. Barozzi et al., "Radiation resistant LGAD design." Nuclear Instruments and Methods in Physics Research Section A: Accelerators, Spectrometers, Detectors and Associated Equipment 919, 16-26 (2019). \href{https://doi.org/10.1016/j.nima.2018.11.121}{DOI: 10.1016/j.nima.2018.11.121}

\bibitem{WU2023167697} K. Wu, X. Jia, T. Yang et al., "Design and testing of LGAD sensor with shallow carbon implantation." Nucl. Instrum. Methods Phys. Res. A 1046, 167697 (2023). \href{https://doi.org/10.1016/j.nima.2022.167697}{DOI: 10.1016/j.nima.2022.167697}

\bibitem{CARTIGLIA201783} N. Cartiglia, A. Staiano, V. Sola et al., "Beam test results of a 16ps timing system based on ultra-fast silicon detectors." Nucl. Instrum. Methods Phys. Res. A 850, 83-88 (2017). \href{https://doi.org/10.1016/j.nima.2017.01.021}{DOI: 10.1016/j.nima.2017.01.021}

\bibitem{GALLRAPP201727} C. Gallrapp, M. Fernández García, S. Hidalgo et al., "Study of gain homogeneity and radiation effects of Low Gain Avalanche Pad Detectors." Nucl. Instrum. Methods Phys. Res. A 875, 27-34 (2017). \href{https://doi.org/10.1016/j.nima.2017.07.038}{DOI: 10.1016/j.nima.2017.07.038}

\bibitem{Medin2022} G. Laštovička-Medin, M. Rebarz, G. Kramberger et al., "Studies of LGAD performance limitations, Single Event Burnout and Gain Suppression, with Femtosecond-Laser and Ion Beams." Nucl. Instrum. Methods Phys. Res. A 1041, 167388 (2022). \href{https://doi.org/10.1016/j.nima.2022.167388}{DOI: 10.1016/j.nima.2022.167388}


\bibitem{Ramo1939}Ramo S., "Currents induced by electron motion Proc." IRE 27 584-5 (1939)
\href{https://doi.org/10.1109/JRPROC.1939.228757}{DOI: 10.1109/JRPROC.1939.228757}

\bibitem{Shockley1938}Shockley W. "Currents to conductors induced by a moving point charge." J. Appl. Phys. 9 635 (1938)
\href{https://doi.org/10.1063/1.1710367}{DOI: 10.1063/1.1710367}

\bibitem{LASTOVICKAMEDIN2022167388}C. Jacoboni, C. Canali, G. Ottaviani et al."A review of some charge transport properties of silicon." Solid-State Electronics, 20,2 (1977),
\href{https://doi.org/10.1016/0038-1101(77)90054-5}{DOI:10.1016/0038-1101(77)90054-5}

\bibitem{Meroli_2011} S. Meroli, D. Passeri, L. Servoli, "Energy loss measurement for charged particles in very thin silicon layers." J. Instrum. 6, P06013 (2011). \href{https://doi.org/10.1088/1748-0221/6/06/P06013}{DOI: 10.1088/1748-0221/6/06/P06013}









\end{thebibliography}

\end{document}